\documentclass[english,aip,prl,graphicx,twocolumn,showpacs]{revtex4}
\usepackage{ae,aecompl}
\usepackage[T1]{fontenc}
\usepackage[latin1]{inputenc}
\usepackage{nicefrac}
\usepackage{float}
\usepackage{ae}
\usepackage{aecompl}
\usepackage{subfigure}
\usepackage{epsfig}
\usepackage{babel}



\begin{document}

\title{The discontinuous nature of the exchange-correlation functional -- critical for strongly correlated systems}

\author{Paula Mori-S\'anchez, Aron J. Cohen and Weitao Yang}

\affiliation{Department of Chemistry, Duke University, Durham, North Carolina
27708, USA}

\begin{abstract}
Standard approximations for the exchange-correlation functional have
been found to give big errors for the linearity condition of
fractional charges, leading to delocalization error, and the
constancy condition of fractional spins, leading to static
correlation error. These two conditions are now unified for states
with both fractional charge and fractional spin: the exact energy
functional is a plane, linear along the fractional charge coordinate
and constant along the fractional spin coordinate with a line of
discontinuity at the integer. This sheds light on the nature of the
derivative discontinuity and calls for explicitly discontinuous
functionals of the density or orbitals that go beyond currently used
smooth approximations. This is key for the application of DFT to
strongly correlated systems.
\end{abstract}

\maketitle

The great success of density functional theory (DFT) \cite{Kohn651133}
is clouded by spectacular failures \cite{Cohen08792} that can manifest themselves in a variety of problems,
from molecular bond stretching to magnetism and the band-gap of materials.
Some of these failures in practical calculations are even considered a breakdown of the theory itself,
as exemplified by the concept of strong-correlation, which often refers to the supposed qualitative collapse
of the single particle picture and hence DFT. In this Letter we present an important condition of the exact DFT
which is missing in currently used approximations and is vital to
understand strongly-correlated systems, in particular their bandgaps.


The fundamental gap of a N-electron system with external potential $v({\bf r})$ is given by the difference of the
ionization energy, $I$, and electron affinity, $A$,
\begin{equation}
E_{\rm gap}^{\rm integer} = \left\{E_v(N-1)-E_v(N)\right\} - \left\{E_v(N) - E_v(N+1)\right\}
\label{intgap}
\end{equation}
or, due to the straight-line behavior of the exact functional \cite{Perdew821691},
by a difference of derivatives at $N$ 
\begin{equation}
E_{\rm gap}^{\rm deriv} = \left\{\left.\frac{\partial E_v}{\partial N}\right|_+
- \left.\frac{\partial E_v}{\partial N}\right|_- \right\}.
\label{derivgap}
\end{equation}

Consider a prototypical example of a strongly correlated system, a stretched H$_2$ molecule. At its
dissociation limit, there are two fractional spin hydrogen
atoms each with half a spin up ($\alpha$) electron and
half a spin down ($\beta$) electron, H$[\textstyle\frac{1}{2},\textstyle\frac{1}{2}]$.
This is a one-electron system that is degenerate in energy \cite{Cohen08121104} with the normal hydrogen atom
with one $\alpha$ electron, H[1,0], and therefore has the same $E_{\rm gap}^{\rm integer}(=0.472E_h)$.
However H$[\textstyle\frac{1}{2},\textstyle\frac{1}{2}]$ has identical $\alpha$ and $\beta$ orbitals by symmetry,
and hence all available functionals give $E_{\rm gap}^{\rm deriv}=0$.
Where is the gap? What is missing from all available exchange-correlation functionals? In other words, what is the
nature of the exact derivative discontinuity \cite{Perdew831884,Sham831888}?

In this Letter we combine fractional charges and fractional spins to
give a unified condition for the exact functional.
This clarifies the behavior of the energy as the number of electrons passes through an integer
and highlights an important discontinuity in the energy expression. An illustrative functional
for the hydrogen atom is developed which clearly shows this discontinuous behavior and moreover
how it manifests itself, with a discontinuous derivative.

We follow the methodology of Yang, Zhang and Ayers \cite{Yang005172}
and examine systems at their dissociation limit. First, we start with
the simplest case. Consider the hydrogen molecular ion H$_{2}^{+}$,
which has one electron, one proton at site $R_{1}$ and another
proton infinitely far away at $R_{2}$. The one electron can be at
either of the two sites, and it can be spin up $(m_{s}=\frac{1}{2})$
or spin down $(m_{s}=-\frac{1}{2})$. Thus there are four degenerate
electronic ground states, $\Psi_{m_{s},l}=\Phi_{m_{s}}\left( R_{l}\right),$
where $\Phi_{m_{s}}\left(R_{l}\right)$ is the ground state of a  hydrogen
atom located at position $R_{l}$ with the spin projection $m_{s}$
and energy $E(H)$. Consider the equally-weighted wavefunction
\begin{equation}
\bar{\Psi}=\sum_{m_{s},l}\Psi_{m_{s},l}/\sqrt{4}.\label{totalwave}\end{equation}
The electron density of this wavefunction is
\begin{equation}
\bar{\rho}=\sum_{l=1}^{2}\rho^{l}=\sum_{l=1}^{2}\sum_{m_{s}}\frac{1}{4}\rho_{m_{s},l},\label{totalRho}\end{equation}
where $\rho_{m_{s},l}$ is the density of $\Psi_{m_{s},l}$. The density
$\bar{\rho}$ of Eq. (\ref{totalRho}) is simply the sum of two identical
densities $\rho^{l}=\frac{1}{4}\left(\rho_{\frac{1}{2},l}+\rho_{-\frac{1}{2},l}\right), l=1,2$,
separated from each other by an infinite distance. The
energy of $\bar{\Psi}$ and $\bar{\rho}$ is $E(H).$ While the density
$\bar{\rho}$ is $v$-representable, $\rho^{l}$ is non-$v$-representable
\cite{Levy821200,Lieb83243}, and is the density of an isolated subsystem
inside the supermolecule described by the wavefunction of Eq. (\ref{totalwave}).


Now we consider the behavior of the exact energy functional, $E_{v}\left[\rho\right]$,
for this density. $E_{v}\left[\rho\right]$ possesses the following
properties:
(A)$E_{v}\left[\rho\right]$ is exact for any (pure-state) $v$-representable
density. Hence, for the total density in Eq. (\ref{totalRho}),
we have
$E_{v}[\bar{\rho}]=E(H)\text{.}$
(B)$E_{v}\left[\rho\right]$ is size extensive. Therefore,
$E_{v}[\bar{\rho}]=\sum_{l=1}^{2}E_{v}[\rho^{l}].$
(C)$E_{v}\left[\rho\right]$ is translationally invariant. Therefore,
$E_{v}[\rho^{1}]=E_{v}[\rho^{2}].$
From (A),(B) and (C) it follows that
\begin{equation}
E_{v}[\rho^{1}]=\textstyle{\frac{1}{2}}E(H).\label{E2degenerate}\end{equation}
Thus, the exact energy for the non-$v$-representable density
$\rho^{1}=\frac{1}{4}(\rho_{\frac{1}{2},1}+\rho_{-\frac{1}{2},1})$
is $\frac{1}{2}E(H)$. This density has a fractional charge of $\frac{1}{2}$
and fractional spins of $\frac{1}{4}$ up-spin and $\frac{1}{4}$
down-spin. Its energy is $\frac{1}{2}E(H)$, the average of hydrogen
atoms with $1$ and $0$ electrons, independent of the fractional spins.
Thus in this specific case, we obtain a condition for the exact functional
for a density with fractional charge and spins, which extends
the previous results of fractional charge \cite{Perdew821691} and fractional spin \cite{Cohen08121104}.

We now generalize our result of Eq. (\ref{E2degenerate}) to include
general fractional charge and fractional spins. We also extend the
discussion to general degeneracies, instead of just a degeneracy
because of spin symmetry. Consider an external potential $v({\bf
\textbf{r}})$ that has two sets of degenerate grounds states:
$N$-electron degenerate ground states with energy $E_{v}(N)$,
wavefunctions $(\Phi_{N,is},i=1,2,...,g_{N})$ and densities
$(\rho_{N,i},i=1,2,...,g_{N})$, and $(N+1)$-electron degenerate
ground states with energy $E_{v}(N+1)$, wavefunctions
$(\Phi_{N+1,j},j=1,2,...,g_{N+1})$ and densities
$(\rho_{N+1,j},j=1,2,...,g_{N+1})$. For the density
$\rho=\frac{1}{q}\sum_{i=1}^{g_{N}}c_{i}\rho_{N,i}+\frac{1}{q}\sum_{j=1}^{g_{N+1}}d_{j}\rho_{N+1,j}$
where $\{c_{i}\}$ and $\{d_{j}\}$ are positive and finite integers,
and satisfy the normalization conditions
$q=\sum_{i=1}^{g_{N}}c_{i}+\sum_{j=1}^{g_{N+1}}d_{j}$,
$p=\sum_{j=1}^{g_{N+1}}d_{j}$, and $q-p=\sum_{i=1}^{g_{N}}c_{i}$,
the exact energy functional satisfies the following equation
\begin{eqnarray}
 & E_{v}\left[\frac{1}{q}\sum_{i=1}^{g_{N}}c_{i}\rho_{N,i}+\frac{1}{q}\sum_{j=1}^{g_{N+1}}d_{j}\rho_{N+1,j}\right]\nonumber \\
= &
\frac{q-p}{q}E_{v}(N)+\frac{p}{q}E_{v}(N+1).\label{Final1}\end{eqnarray}
Eq. (\ref{E2degenerate}) is a special case of the general result,
Eq. (\ref{Final1}), and the proof is given in \cite{PRLSup}. Eq.
(\ref{Final1}) is also valid in first-order reduced density-matrix
functional theory.

We analyze the simple case of a hydrogen atom with general spin
up and spin down occupations, H$[n_\alpha,n_\beta]$, to  illustrate
the scenario of fractional charges and fractional spins combined.
This is of key importance for the consideration of strongly correlated systems
and more especially the band-gap of Mott insulators in the non-magnetic phase.
We focus on the simple strongly correlated system
H$[\textstyle\frac{1}{2},\textstyle\frac{1}{2}]$,
which can be viewed as the infinitely stretched limit of H$_2$ \cite{Cohen08121104}
or as the infinitely expanded
limit of a lattice of hydrogen atoms \cite{Perdew86497} but with
zero spin-density everywhere.
We now address the very interesting question: Is there a gap for H$[\textstyle\frac{1}{2},\textstyle\frac{1}{2}]$?

\begin{figure}[!t]
\includegraphics[width=0.45\textwidth]{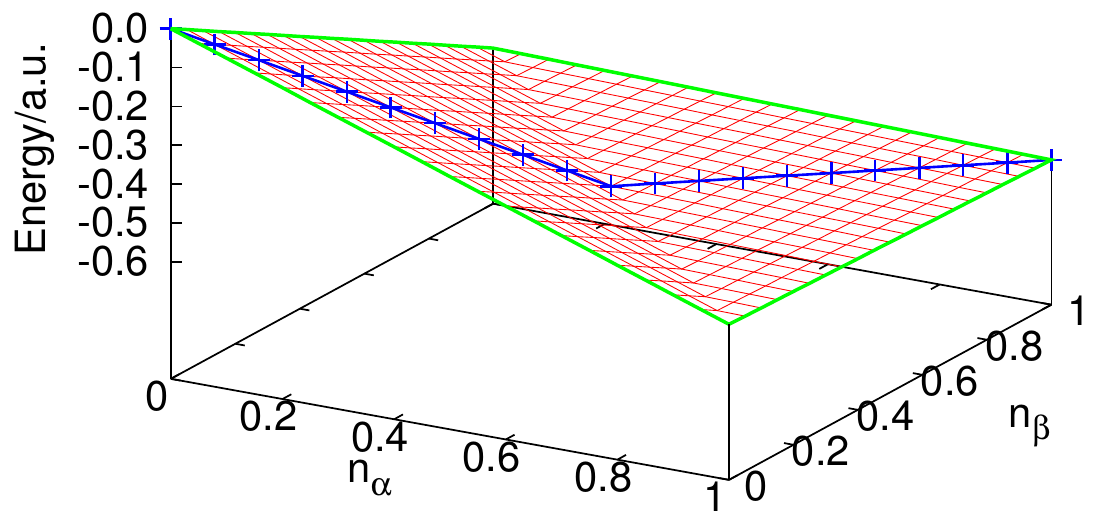}
\caption{Exact energy (a.u.) for densities with fractional charges
and fractional spins in H$[n_\alpha,n_\beta]$.
Fractional charges occur along  the green lines and fractional
spins arise at the intersection $n_\alpha+n_\beta=1$.
The line for $n_\alpha=n_\beta$ is highlighted in blue.}
\end{figure}

The answer to this problem in terms of the total energy can be understood from Eq. (\ref{Final1}).  Fig. 1 shows
the behavior of the exact energy functional
for the hydrogen atom with $0 \le n_\alpha \le 1$,  $0 \le n_\beta \le 1$, and
$0 \le n \le 2$ ($n= n_\alpha + n_\beta$).
The exact functional has
flat plane behavior, linear along the fractional charge coordinate and constant along
the fractional spin coordinate. This gives two flat planes, one for  $0 \le
n \le 1$ and other for  $1 \le n \le 2$, that intersect with a discontinuity at
$n=1$. The simple fractional charge states correspond to the edge lines connecting
$[0,0]$ with $[1,0]$ or $[0,1]$ and then with $[1,1]$, and  the question
of the gap in  H$[\textstyle\frac{1}{2},\textstyle\frac{1}{2}]$
is highlighted by the  blue line  connecting $[0,0]$
with $[\textstyle\frac{1}{2},\textstyle\frac{1}{2}]$ and then
with $[1,1]$. If we now analyze this problem  from a total energy
perspective it is clear that the energy of
H$[\textstyle\frac{1}{2},\textstyle\frac{1}{2}]$ is degenerate
with the normal H$[1,0]$ atom, and also upon addition and removal of
an electron (or any infinitesimal amount of an electron)
the energy change is again exactly the same as the normal
atom. This means that H$[\textstyle\frac{1}{2},\textstyle\frac{1}{2}]$ has an energy
and derivatives, $\left.\frac{\partial E}{\partial N}\right|_\pm$, and therefore gap,
that are exactly the same as H$[1,0]$.

\begin{figure*}[!ht]
\includegraphics[width=0.85\textwidth]{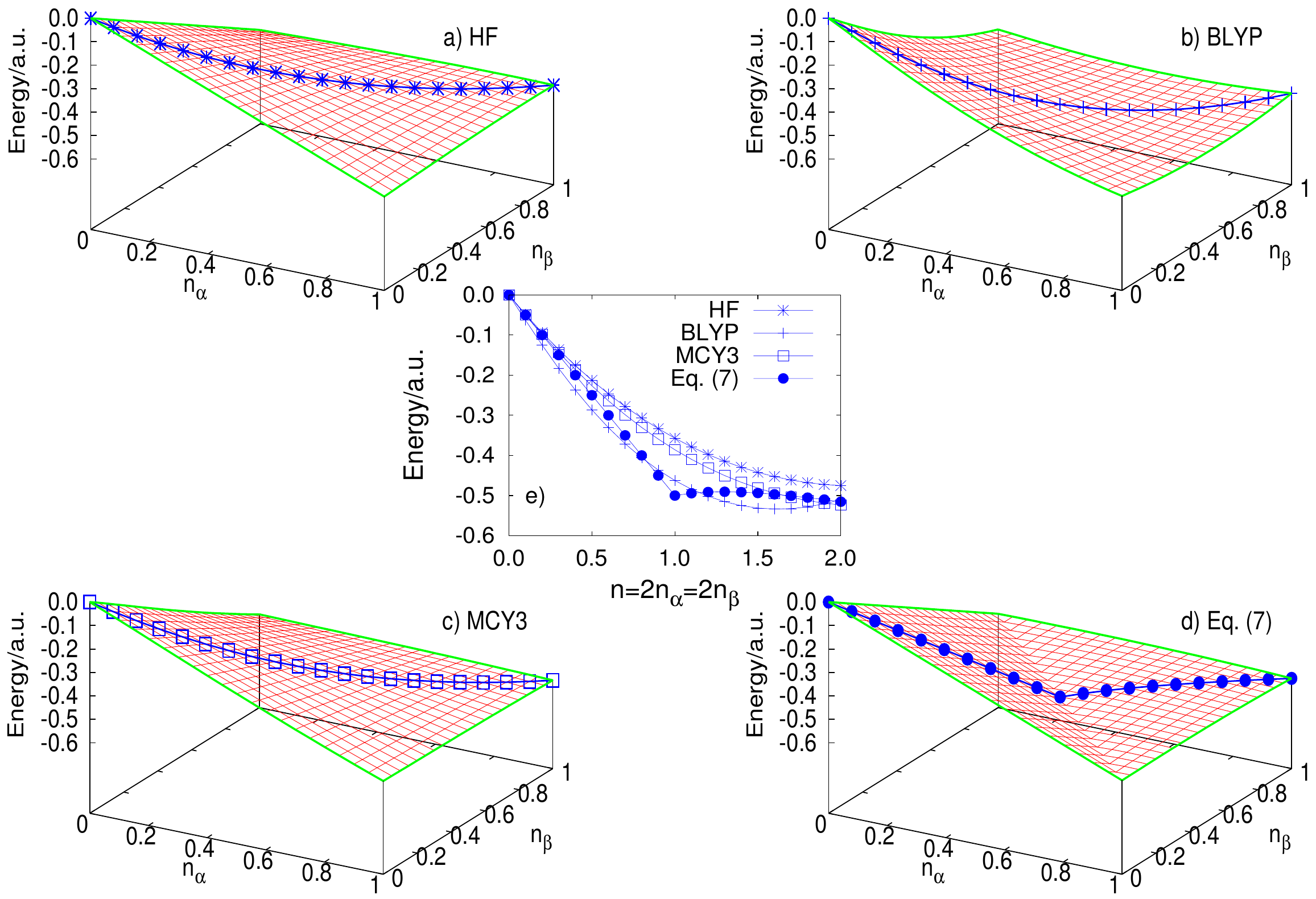}
\caption{The same as Fig. 1 for approximate functionals. All calculations are
self-consistent using a cc-pVQZ basis set.}
\end{figure*}
To gain more insight examine the behavior of several
approximate functionals for the energy of H$[n_\alpha,n_\beta]$ shown in Fig. 2.
If we first consider the fractional charge behavior,
Hartree-Fock (HF) shows the exact straight line behavior between
$[0,0]$ and $[1,0]$ as HF is exact
for one orbital systems. However, it exhibits an incorrect gap for  H$[1,0]$
due to concave curvature for $1\le n\le 2$,
further characterized as localization error \cite{Morisanchez08146401},
and incorrect
energy of H$[1,1]$.
The behavior of semilocal functionals
is exemplified by BLYP in Fig. 2b.  They have an incorrect convex
behavior for fractional charges which usually leads to the
underestimation of $I$ and the overestimation of $A$ from the derivatives \cite{Cohen08115123} and
a delocalization error in larger systems \cite{Morisanchez08146401}.
The MCY3 functional has been developed
to give improved behavior for fractional charges \cite{Cohen07191109},
so it greatly improves the edge lines in Fig. 2c and
hence the gap from the derivatives for H$[1,0]$.
For all of the above functionals there is a failure to correctly predict
the energy of H$[\textstyle\frac{1}{2},\textstyle\frac{1}{2}]$ associated
with the static correlation error of fractional spins \cite{Cohen08121104}.
However much more interesting is the gap of this system.
This is most clearly illustrated in Fig. 2e where only the energy of
the line $n_\alpha=n_\beta$ is shown for $0\le n\le 2$. H$[\textstyle\frac{1}{2},\textstyle\frac{1}{2}]$
is in the middle at $n=1$ and the derivative gap is given by the difference between the slope to the left
and the slope to the right. It is clear that HF, BLYP and MCY3 all have a massive
failure as they incorrectly have no gap for this system. This is a direct consequence of the smooth continuous behavior of
all these different exchange-correlation functionals in terms of the orbitals.

To understand the discontinuous behavior that is needed to give the correct gap,
a simple illustrative functional is built to attempt to describe the whole two plane behavior of
H$[n_\alpha,n_\beta]$ using a linear exchange term, a linear correlation term and the challenging linear
correction of the quadratic Coulomb term;
\begin{widetext}
\begin{equation}
E_{xc}[n_\sigma,\rho_\sigma]  =
\sum_\sigma n_\sigma E_x[\rho_\sigma]
+f_cE_c[\rho_\alpha,\rho_\beta]
 + (n_{\rm high}-n) J[\rho^{n_{\rm low}}_\alpha,\rho^{n_{\rm low}}_\beta]
 + (n-n_{\rm low})J[\rho^{n_{\rm high}}_\alpha,\rho^{n_{\rm high}}_\beta] - J[n_\alpha\rho_\alpha,n_\beta\rho_\beta],
\label{Functional}
\end{equation}
\end{widetext}
with the general repulsion
$J[\rho_a,\rho_b] = \int\int (\rho_a({\bf r})+\rho_b({\bf r})(\rho_a({\bf r}^\prime)+\rho_b({\bf r}^\prime))/
|{\bf r} - {\bf r}^\prime| d {\bf r} d {\bf r}^\prime$.
There is an explicit discontinuity in the correlation and Coulomb correction terms such that
if $n\le 1\ f_c = 0;  n_{\rm low} = 0, \rho^{n_{\rm low}}_\sigma = 0,
n_{\rm high} = 1, \rho^{n_{\rm high}}_\sigma = n_\sigma\rho_{\sigma}/n$
and if
$n\ge1 f_c = (n-1); n_{\rm low} = 1, \rho^{n_{\rm low}}_\sigma = (1-n_{\sigma^\prime})\rho_{\sigma}/(2-n),
n_{\rm high} = 2, \rho^{n_{\rm high}}_\sigma = \rho_\sigma$.
We use HF for  $E_x[\rho_\sigma]$ and LYP for
$E_c[\rho_\alpha,\rho_\beta]$.  Note that irrespective of the form
used for $E_c$ it gives exactly zero correlation energy
for any one-electron system due to the prefactor, $f_c$.
Eq. (\ref{Functional}) is a functional of the orbitals and occupation numbers and
can be viewed within reduced-density-matrix functional theory, which has also been used to tackle strongly correlated systems \cite{Sharma083787},
but in this work we only consider optimization of the orbitals as in standard DFT.
This idea can be generalized and placed much more clearly within DFT if the
normalization of the exact-exchange hole \cite{Becke032971,Perdew07040501}
is used instead of the occupation number.
The performance of this functional in Fig. 2d shows
a qualitative improvement over normal functionals and now resembles the exact
functional of Fig. 1. It recovers the overall feature of two intersecting
planes with a completely flat plane between $0\le n\le 1$ and an almost flat
plane between $1\le n\le 2$, though slightly curved due to the approximate
nature of the dynamic correlation.

To examine the gap of Eq. (\ref{derivgap}), consider the derivatives of the energy expression,
\begin{eqnarray}
\nonumber
E_{\rm gap}^{\rm deriv} = \left.\frac{\partial E_v}{\partial N}\right|_+ - \left.\frac{\partial E_v}{\partial N}\right|_- =& \Delta\epsilon^{\rm KS} + \Delta_{xc} &+ \mathcal{D}_{xc} \\
 = & \Delta\epsilon^{\rm GKS} &+ \mathcal{D}_{xc}
\label{Derivatives}
\end{eqnarray}
where $\Delta\epsilon^{\rm GKS}$ is the difference in
the generalized Kohn-Sham frontier eigenvalues and incorporates all the discontinuity due to a change of orbitals
\cite{Cohen08115123}, including the smooth part, $\Delta_{xc}$, of an orbital dependent exchange-correlation term.
Furthermore $\mathcal{D}_{xc}$ represents only the explicit discontinuity of the exchange-correlation term,
and hence goes beyond previous work \cite{Stadele972089,Cohen08115123}. To understand the nature of the
exact derivative discontinuity \cite{Perdew831884,Sham831888} both $\Delta_{xc}$ and $\mathcal{D}_{xc}$
have to be considered.
The HF, BLYP, MCY3 and all known approximations are only smooth functionals of the orbitals and therefore
$\mathcal{D}_{xc}$ is zero.
Only some methods based on the Hubbard model have a non-zero $\mathcal{D}_{xc}$ \cite{Lima02601}.
The new functional, Eq. (\ref{Functional}) has an explicit derivative discontinuity at $n=1$
\begin{equation}
\mathcal{D}_{xc}^{\rm Eq. (\ref{Functional})} = -2J[\rho^1_\alpha,\rho^1_\beta] + J[\rho^2_\alpha,\rho^2_\beta] + E_c^{\rm LYP}[\rho^2_\alpha,\rho^2_\beta].
\end{equation}

\begin{table}[!t]
\caption{Energy gap (a.u.) from the derivatives (Eq. \ref{Derivatives}).}
\begin{tabular}{lcccccc}
\hline\hline
                                &     &  BLYP  &  HF    & MCY3   & Eq. (\ref{Functional})   &  Expt. \\
\hline
H[1,0]                          &$I-A$&  0.250 & 0.546  &  0.425 & 0.576 &  0.472 \\
                                &$I$  &  0.272 & 0.500  &  0.448 & 0.500 &  0.500 \\
                                &$A$  &  0.022 & -0.046 &  0.023 & -0.076&  0.028 \\
\hline
H[$\frac{1}{2},\frac{1}{2}$]    &$I-A$&   0    & 0      & 0      & 0.576 &  0.472\\
                                &$I$  &  0.239 & 0.227  & 0.252  & 0.500 &  0.500\\
                                &$A$  &  0.239 & 0.227  & 0.252  & -0.076&  0.028\\
\hline\hline
\end{tabular}
\end{table}

Table 1 shows $I$, $A$ and the gap from the derivatives.
The gap of H$[1,0]$ is underpredicted by BLYP due
to its convex behavior and overpredicted by HF due to its
concave behavior and bad endpoint, while MCY3 improves in this case due to its straight line behavior.
However none of these normal functionals offer a gap for H$[\textstyle\frac{1}{2},\textstyle\frac{1}{2}]$ due to
a lack of $\mathcal{D}_{xc}$.
The functional in Eq. (\ref{Functional})
reveals the correct picture with a reasonable gap that is the
same for both H$[\textstyle\frac{1}{2},\textstyle\frac{1}{2}]$
and H[1,0], although $A$ is slightly underpredicted due to the
concave nature for $1\le n\le 2$.  Moreover these two gaps are both due
to the explicit discontinuity of the functional, $\mathcal{D}_{xc}$, and there
is no contribution from the orbital discontinuity, as
the $\alpha$ and $\beta$ orbitals from the self-consistent solution of Eq. (\ref{Functional}) are always degenerate at $n=1$.
This raises an interesting question about normal functionals
which erroneously break this degeneracy for $n=1$ when $n_\alpha \ne n_\beta$
They give a gap that is therefore of the wrong nature, due to a discontinuity because of the
orbital dependence, which only mimics the correct explicit discontinuity with degenerate orbitals.
This erroneous behavior of normal functionals must no doubt contribute to the incorrect prediction of quantities
related to these orbitals such as hyperfine, spin and magnetic properties.

In this Letter a new exact condition for the energy functional is derived that shows the combined behavior
for fractional charge and spins. It shows a discontinuous behavior when passing through the integer that
reveals the explicitly discontinuous nature of the derivative of the exchange-correlation functional.
This is most clearly highlighted by the gap of H$[\textstyle\frac{1}{2},\textstyle\frac{1}{2}]$.
A gap for this system only appears if the exchange-correlation functional has an explicit discontinuity,
because the orbitals are degenerate and give no contribution.
For Mott insulators, the unit cell has fractional spins corresponding to different magnetic phases
and the correct gap prediction critically depends on this explicit discontinuity.
For the future it is crucial to develop better approximations that go beyond smooth orbital functionals
by including an explicit discontinuity, along the line suggested by Eq. (\ref{Functional}).
This is essential for the advancement of DFT towards the calculation of strongly correlated systems.

Support from NSF is greatly appreciated.

\appendix
\section{Appendix: Proof of Eq. (6)}

We follow the methodology of Yang, Zhang and Ayers
\cite{Yang005172}. Consider a supramolecular system with the
following external potential $v_{\rm total}({\bf
r})=\sum_{l=1}^{q}v(\mathbf{r}-\mathbf{R}_{l})$; namely it has $q$
copies of the potential $v({\bf \textbf{r}})$ each located at a site
${\bf \textbf{R}}_{l}$ which are infinitely far away from each
other. There is a total of $qN+p$ electrons ($N$, $p$ and $q$ are
all positive and finite integers and $q>p$). Since the sites are
separated by infinite distances, the total system is simply composed
of $q$ subsystems in identical external potentials $v({\bf
\textbf{r}})$, with no interaction between the subsystems. Its
ground state has $(q-p)$ $N$-electron subsystems and $p$
$(N+1)$-electron subsystems, assuming the convexity condition
\begin{equation}
E_{v}(N)\leq(E_{v}(N+1)+E_{v}(N-1))/2,\label{convex0}\end{equation}
which is known to hold for atoms and molecules from experimental
data \cite{Parr89,Perdew821691}. The ground state is degenerate and
its energy is \begin{equation}
(q-p)E_{v}(N)+pE_{v}(N+1)\label{TotalE_supra}\end{equation}

The total ground-state wavefunction is an antisymmetric product of
$q$ separated ground state wavefunctions. One possible state is
following: The first $p$ locations, ${\bf \textbf{R}}_{1}...{\bf
\textbf{R}}_{p},$ each have $(N+1)$ electrons; within these $p$
locations, the first $d_{1}$ sites have the degenerate wavefunction
$\Phi_{N+1,1}$, the second $d_{2}$ sites have the degenerate
wavefunction $\Phi_{N+1,2}$, ..., and the last $d_{g_{N+1}}$ sites
have the degenerate wavefunction $\Phi_{N+1,g_{N+1}}$. In this way,
$p=\sum_{j=1}^{g_{N+1}}d_{j}$. The remaining $q-p$ locations, ${\bf
\textbf{R}}_{p+1}...{\bf \textbf{R}}_{q},$ each have $N$ electrons;
within these $q-p$ locations, the first $c_{1}$ sites have the
degenerate wavefunction $\Phi_{N,1}$, the second $c_{2}$ sites have
the degenerate wavefunction $\Phi_{N,2}$, ... and the last
$c_{g_{N}}$ sites have the degenerate wavefunction $\Phi_{N,g_{N}}$.
In this way, $q-p=\sum_{i=1}^{g_{N}}c_{i}$. Then this state has the
wave function
\begin{eqnarray}
\Psi_{1} & = & \hat{A}\{\Phi_{N+1,1}\left({\bf \textbf{R}}_{1}\right)...\Phi_{N+1,1}\left({\bf \textbf{R}}_{d_{1}}\right)\nonumber \\
 &  & \Phi_{N+1,2}\left({\bf {\bf \textbf{R}}_{d_{1}+1}}\right)...\Phi_{N+1,2}\left({\bf \textbf{R}}_{d_{1}+d_{2}}\right)\nonumber \\
 &  & ...\nonumber \\
 &  & \Phi_{N,1}\left({\bf \textbf{R}}_{p+1}\right)...\Phi_{N,1}\left({\bf \textbf{R}}_{p+c_{1}}\right)\nonumber \\
 &  & \Phi_{N,2}\left({\bf \textbf{R}}_{p+c_{1}+1}\right)...\Phi_{N+1,2}\left({\bf \textbf{R}}_{p+c_{1}+c_{2}}\right)...\}\label{eq:Psi1}
\end{eqnarray}
Permutation of any two locations with different states
($\Phi_{N,i},$ or $\Phi_{N+1,j}$) generates a different
$(qN+p)$-electron wavefunction. There are a total of
$m=\nicefrac{q!}{\left(\prod_{i}^{g_{N}}c_{i}!\prod_{j}^{g_{N+1}}d_{j}!\right)}$
such degenerate wavefunctions.

For any wave function $\Psi_{k},$ a particular site ${\bf
\textbf{R}}_{l}$ can either have the wavefunction $\Phi_{N,i},$ or
$\Phi_{N+1,j}$. In all such wavefunctions $\{\Psi_{k},k=1,..,m\}$,
the number of times any location ${\bf R}_{s}$ having the wave
function $\Phi_{N+1,n}$ is equal to
$m_{N+1,n}=\nicefrac{(q-1)!}{\left((c_{n}-1)!\prod_{i\neq
n}^{g_{N}}c_{i}!\prod_{j}^{g_{N+1}}d_{j}!\right)}=\nicefrac{mc_{n}}{q}$
and the corresponding number for $\Phi_{N,n}$ is equal to
$m_{N,n}=\nicefrac{(q-1)!}{\left((d_{n}-1)!\prod_{i}^{g_{N}}c_{i}!\prod_{j\neq
n}^{g_{N+1}}d_{j}!\right)}=\nicefrac{md_{n}}{q}$. In analogy to Eq.
(3) of the paper, the following equally-weighted wavefunction is
also a degenerate wavefunction
\begin{equation}
\bar{\Psi}=\frac{1}{\sqrt{m}}\sum_{k=1}^{m}\Psi_{k},\label{totalwave2}\end{equation}
 the density of which is
\begin{eqnarray}
\bar{\rho} & = &
\sum_{l=1}^{q}\frac{1}{q}\left(\sum_{i=1}^{g_{N}}c_{i}\rho_{N,i}(\mathbf{R}_{l})+\sum_{j=1}^{g_{N+1}}d_{j}\rho_{N+1,j}(\mathbf{R}_{l})\right).\label{totalRho2}
\end{eqnarray}
In this particular state, with the degenerate energy of Eq.
(\ref{TotalE_supra}), all the $q$ subsystems have the same electron
density except by translation. Following the arguments leading to
(A), (B) and (C) of the paper we have
\begin{eqnarray}
  &E_{v}\left[\frac{1}{q}\sum_{i=1}^{g_{N}}c_{i}\rho_{N,i}(\mathbf{R}_{l})+\frac{1}{q}\sum_{j=1}^{g_{N+1}}d_{j}\rho_{N+1,j}(\mathbf{R}_{l})\right]\nonumber \\
=&
\frac{q-p}{q}E_{v}(N)+\frac{p}{q}E_{v}(N+1)\label{E2degenerate2}\end{eqnarray}
 which is just Eq. (6) of the paper, for the site $\mathbf{R}_{l}$.

\end{document}